\documentclass[12pt]{article}
\usepackage{graphicx}
\usepackage{amssymb,amsmath}

\newcommand\pubnumber{DPF2013-59}
\newcommand\pubdate{\today}

\def\napoli{Department of Physics\\
Tallahassee, Florida 32306-4350}
\def\support{\footnote{Work supported by DE-SC0010102, Dept of Energy.}}

\def\Title#1{\begin{center} {\Large #1 } \end{center}}
\def\Author#1{\begin{center}{ \sc #1} \end{center}}
\def\Address#1{\begin{center}{ \it #1} \end{center}}

\newcommand\pubblock{\rightline{\begin{tabular}{l} \pubnumber\\
         \pubdate  \end{tabular}}}
\newenvironment{Abstract}{\begin{quotation}  }{\end{quotation}}
\newenvironment{Presented}{\begin{quotation} \begin{center} 
             PRESENTED AT\end{center}\bigskip 
      \begin{center}\begin{large}}{\end{large}\end{center} \end{quotation}}

\def\beq{\begin{equation}}
\def\eeq#1{\label{#1}\end{equation}}
\def\eeqn{\end{equation}}

\def\beqa{\begin{eqnarray}}
\def\eeqa#1{\label{#1}\end{eqnarray}}
\def\eeqan{\end{eqnarray}}

\let\bar=\overbar

\def\Dslash{\not{\hbox{\kern-4pt $D$}}}
\def\dslash{\not{\hbox{\kern-2pt $\del$}}}

\def\msb{{\bar{\ssstyle M \kern -1pt S}}}

\begin{document}
\begin{titlepage}
\pubblock

\vfill
\Title{Threshold resummation in direct photon production 
		and its implications on the large-x gluon PDF}
\vfill
\Author{Nobuo Sato\support}
\def\emailpj{\footnote{nsato@hep.fsu.edu}}
\Address{\napoli}
\vfill
\begin{Abstract}
Currently, the gluon distribution function is mainly constrained by jet data. 
Yet, its high-x behaviour is largely unknown.
This kinematic region is important, for instance, for the  understanding
of the production of a massive state at forward rapidities at the LHC. 
In the past, single inclusive direct photon data from  
fixed target experiments was used to constrain the gluon distribution due to 
its dominant contribution from the $qg \rightarrow\gamma q$ subprocess in 
proton-proton collisions. 
Due to disagreement with its theoretical
predictions at next-to-leading order (NLO) in perturbative QCD (pQCD), 
direct photon data have been excluded from global fits. 
This talk will discuss an improvement to such theoretical predictions
by including \emph{threshold resummation} at next-to-leading logarithmic 
accuracy (NLL) and its impact on gluon distribution 
at large-x using Bayesian reweighting technique.
\end{Abstract}
\vfill
\begin{Presented}
DPF 2013\\
The Meeting of the American Physical Society\\
Division of Particles and Fields\\
Santa Cruz, California, August 13--17, 2013\\
\end{Presented}	
\vfill
\end{titlepage}
\def\thefootnote{\fnsymbol{footnote}}
\setcounter{footnote}{0}
%
\section{Introduction}
The theoretical predictions, of any process at hadron colliders depends
on our knowledge of parton distribution functions (PDFs).
These functions are currently not computable from first principles but 
instead they are extracted from data. 
Such task, commonly known as \emph{global fits}, 
has been conducted by several collaborations that
provide numerical tables for the  
distributions at different kinematial regions including central values
as well as  uncertainties that  behave differently among 
the various parton species.
In particular, the gluon distribution is unconstrained at high-$x$. 

Historically, the information on high-$x$ gluon PDF
was originally extracted from the data of  single inclusive 
direct photon production at fixed target experiments. 
However, these data have largely been excluded from global fits
due to disagreement with its theoretical description at NLO in pQCD. 
As today, the inclusive jet data from Tevatron experiment 
provides most of the information on the gluon PDF. 
Yet, its uncertainties at high-$x$ are still significantly large, so that 
any prediction of observables that are sensitive to the gluon PDF in this
region are not reliable.   
More recently it was shown in Ref.~\cite{d'Enterria:2012yj}
that isolated photon production from collider data can be consistently 
included in global fits and provides constraints on the 
gluon PDF in the region $x\sim0.3$.

The disagreement between fixed target direct photon data and
its theoretical calculation might be alleviated by improving the 
calculation beyond the NLO. 
One kind of improvement is called \emph{threshold resummation}.
Beyond the leading order in pQCD, near partonic threshold when the initial 
partons have just the enough energy to produce the high-$p_T$ photon, 
the phase space for extra gluon radiation is limited.
As a consequence, the infrared cancellations between the virtual and 
real contributions are not exact, leaving logs of the form 
$\alpha_s^k\ln^{2k}(1-z)$ where $z=1$ is the partonic threshold. 
The resummation of these threshold logs has been 
described  in Ref.~\cite{Catani:1998tm,deFlorian:2005wf} 
at NLL+NLO. 

In this paper we present preliminary results on the potential impact 
of direct photon data from fixed target experiments on gluon PDF using 
the theoretical improvement mentioned above.
To quantify such impact, we use a reweighting technique,
based on Bayesian inference, following the ideas from Ref.~\cite{Giele:1998gw,Watt:2012tq}.
The document is organized as follow: in the next section 
a summary of the reweighting method is presented.
A discussion of the preliminary results are given in 
Sec.~\ref{sec:preliminary}.
We present the conclusions in Sec.~\ref{sec:conclusion}.

\section{The reweighting method}
\label{sec:reweighting}
We start by constructing $N=100$ random PDFs distributed according to 
\begin{align}
f_k(x)=f_0(x)+\sum_j (f^{\pm}_j(x)-f_0(x))|R_{kj}|,
\label{eq:mc}
\end{align}
where $f$ labels the various parton species.
We will refer to members of the set $\{f_k\}$ as \emph{replicas}. 
For simplicity we have dropped the implicit dependence of the PDFs on 
the factorization scale.  
The central values are given by $f_{0}$ and the confidence interval is
encoded in the quantities $f_j^{\pm}$ (See for instance \cite{Watt:2012tq}).
$R_{kj}$ are normally distributed random numbers with mean zero and 
variance one. The $\pm$ in $f_j^{\pm}$ is chosen according to the sign of
$R_{kj}$. 
For each replica, we compute the theoretical predictions for the direct
photon data and the corresponding $\chi^2_k$ using the $t_0$-method 
developed in Ref.~\cite{Ball:2009qv}.
Then, we assign \emph{weights}  of the form
\begin{align}
w_k\propto \exp\left(-\frac{1}{2}\chi^2_k\right),
\end{align}
to each replica.
The weights are normalized by demanding $N=\sum_k w_k$. 
The impact of direct photon data on gluon PDF can be quantified by 
computing the weighted expectation values and variances:
\begin{align}
\text{E}[f(x)]&=\frac{1}{N}\sum_k w_k\,f_k(x)\notag\\
\text{Var}[f(x)]&=\frac{1}{N}\sum_k w_k\,(f_k(x)-\text{E}[f(x)])^2.
\end{align}
The corresponding quantities without reweighting 
(or equivalently \emph{unweighted}) are obtained  
by setting all the weights equal to one.

\section{Preliminary results}
\label{sec:preliminary}

\begin{figure}[htb]
\centering
\includegraphics[width=\textwidth]{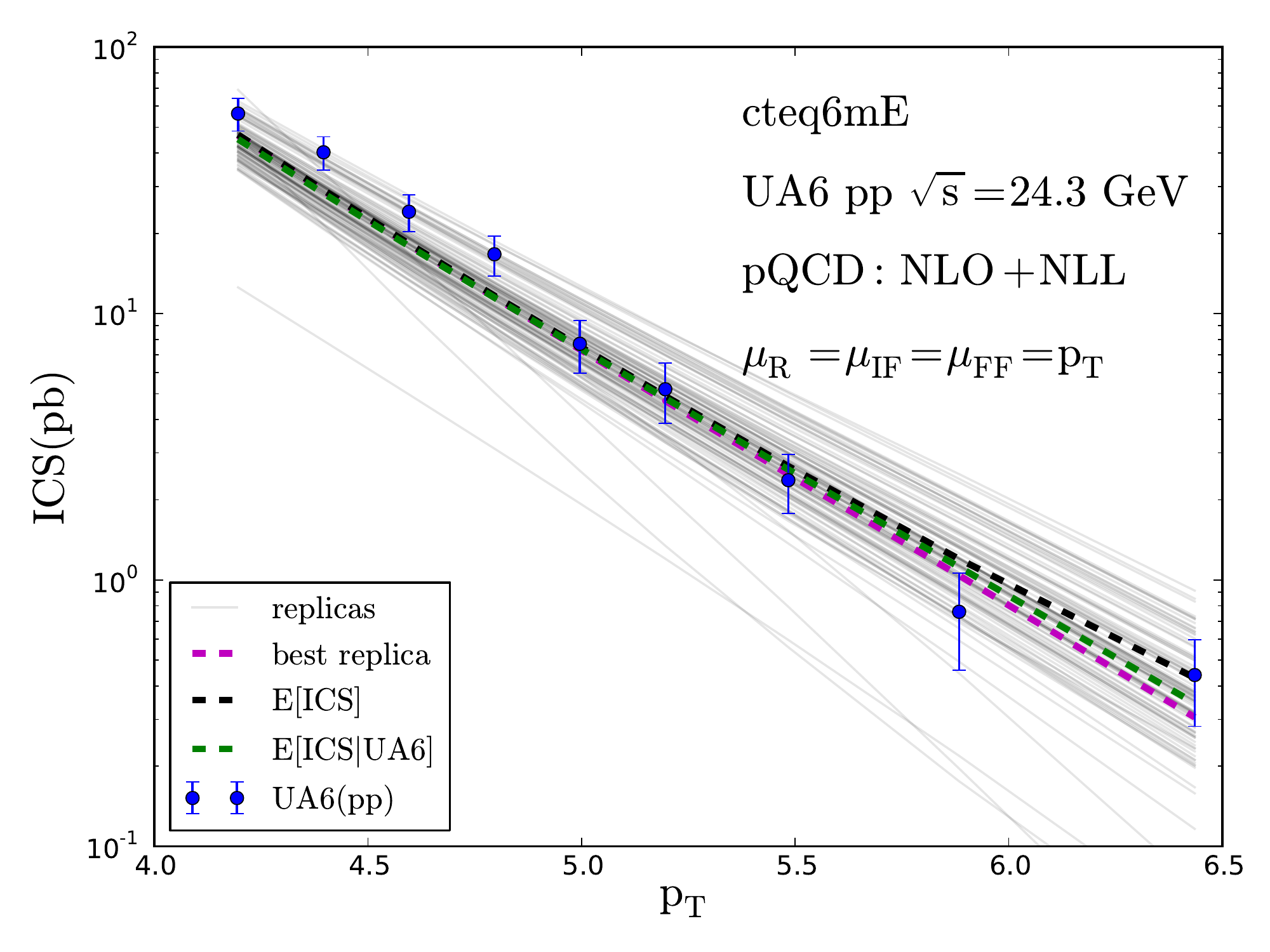}
\caption{Invariant cross section (ICS) vs. $p_t$. 
The grey curves are the predictions from the replicas.
The best replica (the one with the highest weight) is
plotted as dashed magenta curve.
$\text{E}[\text{ICS}]$ and $\text{E}[\text{ICS}|\text{UA6}]$
are the expectation values for the unweighted and weighted replicas
respectively.
}
\label{fig:ICS}
\end{figure}

Here, we present the impact of the UA6 experiment 
($pp\rightarrow\gamma+X$ $\sqrt{s}=24.3$ GeV)
~\cite{Ballocchi:1998au} on the Cteq6m gluon PDF~\cite{Pumplin:2002vw}.
In Fig.~\ref{fig:ICS}, we have plotted the invariant cross section (ICS)
as a function of $p_T$.
The plot also includes the theory predictions from the replicas
and the corresponding unweighted and weighted 
expectation values labeled as $\text{E}[\text{ICS}]$ and 
$\text{E}[\text{ICS}|\text{UA6}]$ respectively.
The renormalization and factorization scales have been set equal to $p_T$.
The fact that the expectation values are close to each other implies 
a good agreement between the UA6 data and its 
theoretical predictions at NLO+NLL.

\begin{figure}[!htb]
\centering
\includegraphics[width=\textwidth]{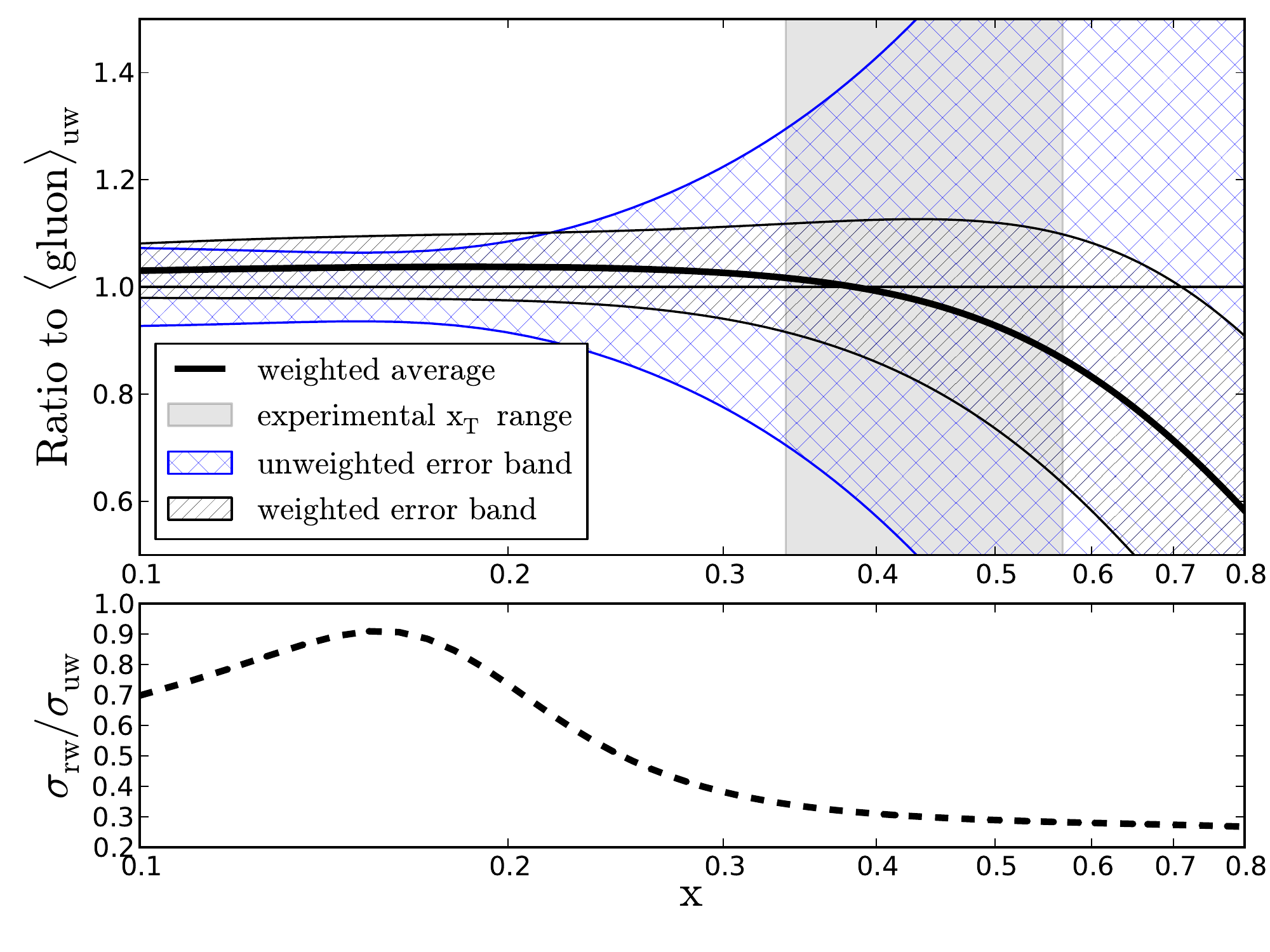}
\caption{Top: Gluon PDF uncertainty bands normalized to the expectation 
value of unweighted gluon PDF.
The grey band corresponds to the region of UA6 direct photon data.
Bottom: ratio of reweighted variance ($\sigma_{rw}$) over unweigted
variance ($\sigma_{uw}$). 
 }
\label{fig:ratio}
\end{figure}

\begin{figure}[!htb]
\centering
\includegraphics[width=\textwidth]{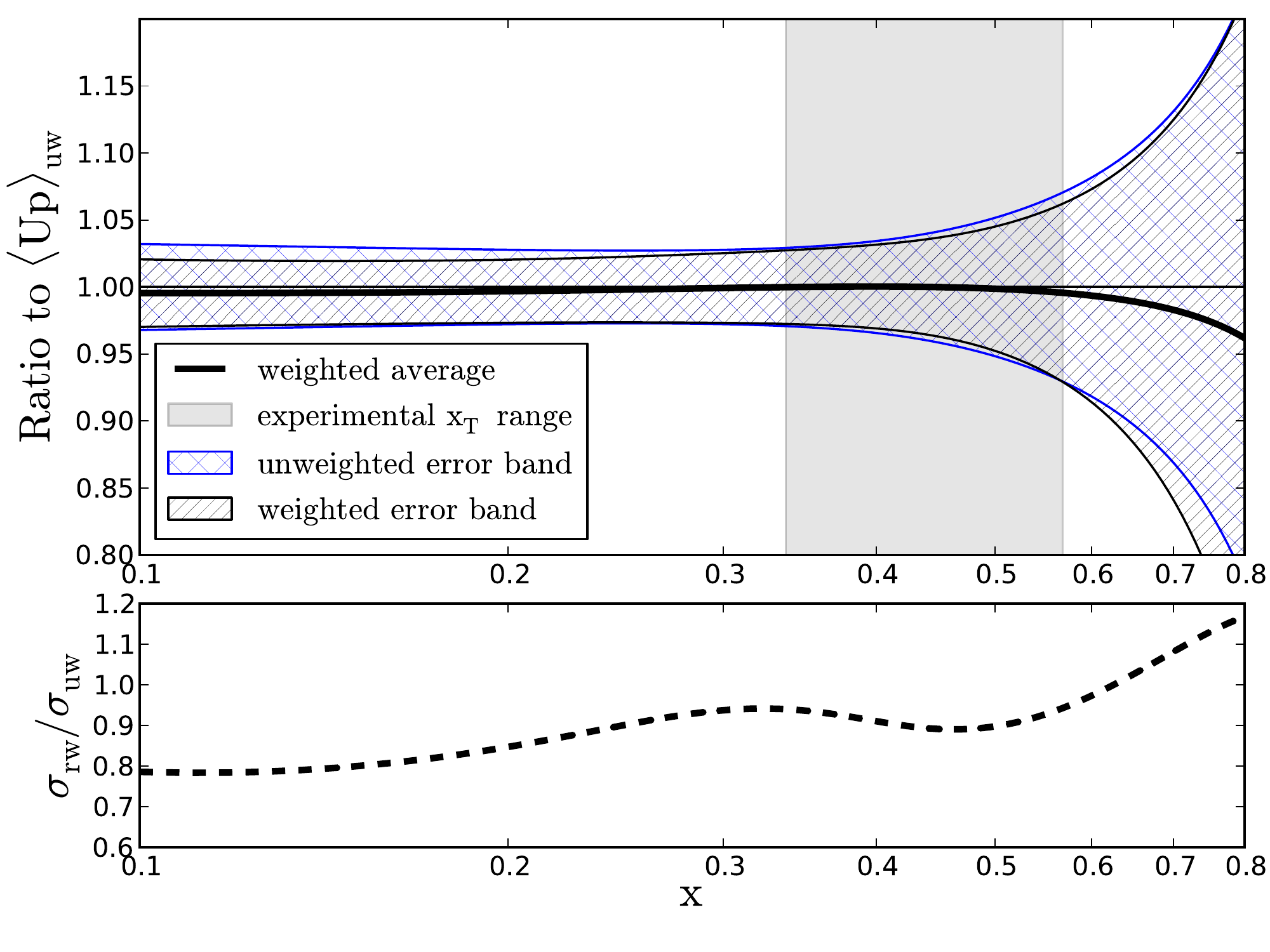}
\caption{Same as Fig.~\ref{fig:ratio} for Up quark PDF. }
\label{fig:ratioup}
\end{figure}

In Fig.~\ref{fig:ratio} we present the gluon PDF uncertainty 
band (before and after the reweighting) at $Q=10$ GeV as a function of $x$.
In both cases we normalized respect to the expectation value from the 
unweighted results.
Clearly a significant reduction in the uncertainty band is observed 
due to the presence of the UA6 data (the kinematic coverage is shown in 
the plot as a shaded region).
The sensitivity of gluon PDF with direct photon data is expected since 
its production in proton-proton collisions is dominated 
by the $qg\rightarrow\gamma q$ subprocess. 
In Fig.~\ref{fig:ratioup} we have plotted a similar figure  as 
Fig.~\ref{fig:ratio} for the up quark distribution.
In this case, since the up quark distribution is relatively well 
constrained in the kinematic region of UA6 data, the latter does not
provide new information on the up quark PDF. 
%

\section{Conclusion}
\label{sec:conclusion}
We have presented a preliminary results on Cteq6m gluon PDF reweighting 
using the data of single inclusive direct photon production from  
the UA6 experiment.
The theoretical predictions for the data has been computed using 
threshold resummation at NLL+NLO in  pQCD. 
After reweighting, a significant reduction in gluon PDF uncertainty
is observed. 
Further analysis exploring other PDFs sets will be presented in an
a future work (in preparation).

\end{document}